\def\prd{Phys. Rev. D}
\def\prl{Phys. Rev. Lett.}

\def\apjs{Astrophys. J.Suppl.}

\def\aap{Astr. Astrophys.}

\def\jcap{JCAP}

\def\procspie{Proc. SPIE}

\def \<{\langle}
\def \>{\rangle}

\newcommand{\Abs}{\abstract}
\newcommand{\Ack}{\acknowledgments}
\newcommand{\mktt}{\maketitle}

\documentclass[a4paper,11pt]{article}
\pdfoutput=1

\def\orcid#1{\kern .08em\href{https://orcid.org/#1}{\includegraphics[width=1.0em]{orcid.jpg}}}

\usepackage{jcappub,graphicx,epsfig,natbib,color,times,bm,amsmath,multirow,amssymb,mathtools,mathrsfs,extarrows}
\usepackage{xcolor, academicons, booktabs}
\usepackage[ddmmyyyy,hhmmss]{datetime}

\begin{document}

\title{Removal of point source leakage from time-order data filtering}



\author[b,a]{Zhaoxuan Zhang,}
\author[b]{Lu Huang,}
\author[c]{Yang Liu,}
\author[c]{Si-Yu Li,}
\author[b,c,d]{Le Zhang,}
\author[a,c]{Hao Liu}\emailAdd{ustc\_liuhao@163.com}

\affiliation[a]{School of Physics and optoelectronics engineering, Anhui University, 111 Jiulong Road, Hefei, Anhui, 230601, China}

\affiliation[b]{School of Physics and Astronomy, Sun Yat-sen University, 2 Daxue Road, Tangjia, Zhuhai, 519082, People's Republic of China}

\affiliation[c]{Key Laboratory of Particle and Astrophysics, Institute of High Energy Physics, CAS, 19B YuQuan Road, Beijing, 100049, China }

\affiliation[d]{CSST Science Center for the Guangdong-Hong Kong-Macau Greater Bay Area, Zhuhai 519082, People's Republic of China}

\affiliation[e]{Peng Cheng Laboratory, No.2, Xingke 1st Street, Shenzhen 518000, People’s Republic of China}

\Abs{
Time-ordered data (TOD) from ground-based CMB experiments are generally filtered before map-making to remove or reduce the contamination from the ground and the atmospheric emissions. However, when the observation region contains strong point sources, the filtering process will result in considerable leakage around the point sources in a measured CMB map, and leave spurious polarization signals. Therefore, such signals need to be assessed and removed before CMB science exploitation. In this work, we present a new method that we call "template fitting" and can effectively remove these leakage signals in pixel
domain, not only satisfying the requirement for measuring primordial gravitational 
waves from CMB-$B$ modes, but also avoiding time-consuming operations on TOD.
}

\mktt
\tableofcontents
\section{Introduction}\label{sec:intro}
The cosmic microwave background (CMB) temperature and polarization
anisotropies are strong observational evidence of the inflationary
expansion history of the Universe. Especially, the detection of $B$-mode
CMB polarization in form of the tensor to scalar ratio $r$ is crucial
for confirming the existence of the gravitational waves in the early
Universe, which is a natural consequence of the inflationary potential.
Several space and ground-based experiments are devoted to constraining
$r$, including the BICEP series \citep{2003SPIE.4843..284K}, Planck
\citep{2006astro.ph..4069T}, QUIJOTE \citep{2010ASSP...14..127R}, ACTPol
\citep{2010SPIE.7741E..1SN}, SPTPol \citep{2012SPIE.8452E..1EA}. Current
observations already limit the tensor to scalar ratio to $r\le
0.036-0.1$ \citep{2013ApJS..208...19H, 2018arXiv180706209P,
2015PhRvL.114j1301B, 2018PhRvL.121v1301B, PhysRevLett.127.151301}, and
the forthcoming experiments including POLARBEAR
\citep{2011arXiv1110.2101K}, LiteBIRD \citep{2012SPIE.8442E..19H},
CMB-S4 \citep{2016arXiv161002743A}, the Simons Observatory
\citep{2019JCAP...02..056A}, and AliCPT \citep{doi.10.1093.nsr.nwy019}
will devote to reaching a sensitivity of $r=0.001$. This inevitably
requires dedicated treatments of all kinds of contamination and
systematics. Especially, all available CMB experiments in the next
$5\sim10$ years are ground-based, which are ineluctably contaminated by
the atmosphere and ground emissions. Therefore, their time-order data
(TOD) needs to be filtered to alleviate these contaminations.

Usually, in a ground-based experiment (e.g., BICEP or AliCPT), the
observed TOD is converted to the T, Q, and U maps by the data analysis
pipeline that contains data splitting and cutting, pointing and
polarization orientation reconstruction (for each detector), time domain
filtering, and the final time-to-pixel domain map making. First of all,
the TOD are split in units of ``halfscan'', and ``scanset'' (tens of
neighbouring halfscans), and bad data are removed at certain thresholds.
After that, the pointing trajectory and polarization orientation for
each detector are constructed from the encoder data, GPS time, site
location as well as the focal plane structure, then the TOD are
high-passed to suppress long-distance correlations arising from noise
sources or systematic errors. In order to remove the atmospheric
radiation present in the data, a polynomial filter (typically of the 3rd order) is applied, and to handle noise associated with the ground
coordinate system, such as ground reflections/emissions, templates are
constructed and removed for each scanset, which is called a ground
subtraction filter. In addition, as a polarization experiment targeting
the CMB $B$-mode, the temperature-polarization leakages need to be removed
as much as possible, e.g., by using a de-projection filter on scansets
to suppress the leakage due to the beam mismatch of orthogonal
polarization detector pairs. It is worth noting that all the filters
mentioned above are linear, so it is possible to implement them
either as direct time domain operations, or as pixel domain matrix
operations, which are completely equivalent. After all these operations,
the TOD is weighted by the inverse variance estimated from each scanset,
and then projected and co-added on the sky for each map pixel, to
produce the final observed map.

Unfortunately, although the TOD filtering can efficiently remove the
atmosphere/ground emissions, as well as fixing some other errors, it
will also remove part of the CMB signal and cause leakages from point
sources to the pixel domain regions around them. Because the latter
usually has no preference for the $E$- and $B$-modes, when the point source is
strong, it can significantly contaminate the weak primordial $B$-mode
signal in a large pixel domain region. For measuring the primordial gravitational waves through the extremely faint $B$-mode signal, this kind of leakage has to be
removed accurately.

In principle, if one knows the positions of point sources, then it is
possible to identify their locations in TOD and cut the corresponding
TOD segments to prevent the leakage due to filtered point sources. However, this
will cause several problems: 1) Not all point sources are strong enough
to be identified from the TOD, because the TOD is much noisier than
the final stacked sky map. 2) Removal of the TOD segments with the point
source will break the integrity of the TOD and cause inconvenience and issues in filtering and map-making. 3) Typically, operations on TOD are  
very time-consuming, so removal of the point source leakage directly in the TOD becomes in feasible in practice.

In this work, we introduce a new method to remove the point source leakage due to filtering of the TOD. This method is based on ~\citep{2018arXiv181104691L} and operates mainly in the pixel domain.
The main idea is to construct ideal and realistic templates of the
leakages in the pixel domain, and then remove the leakages by linear
regression. The advantage of this method is obvious: 1) This method will
not cut the TOD, which is safe, convenient and friendly to all types of
TOD operations. 2) From the test results, this method can successfully remove the point source leakage down to the level satisfying the detection requirement of $r\le 0.001$. 3) This method operates mainly in the pixel
domain, which is fast and easy to implement.

The structure of this paper is as follows: in Sect. \ref{sec:methods}, we
introduce our removal of point source leakage method for both single and
multiple point sources. We give examples for these two situations and
use actual point sources data to verify our method in Sect.
\ref{sec:examples}. Finally, we summarize and conclude in Sect.~\ref{sec:disscuss}.

\section{Methods}\label{sec:methods}
 The core concept to alleviate the point source leakages on the final
    sky map is based on the fact, that all TOD operations and their
    effects in the pixel domain sky maps are linear. In order to remove
    the leakages by linear regression, the fundamental operation is to
    create pixel domain leakage templates. Since the data we obtain from
    the pipeline is always filtered, two types of templates can be
    produced: ideal and realistic. The main difference between them is
    that the ideal template requires  complete knowledge of the beam
    profile, which is usually unavailable\footnote{ A Gaussian beam
    profile is often used as an approximation.}; whereas a realistic
    template is constructed directly from the product of the pipeline,
    which is always available. Certainly, the performance of the ideal
    template is better than the realistic template, as we will mention below
    that the results of cleaning by the realistic template are also
    acceptable.

    \subsection{The ideal template}\label{sub:ideal template}

    Construction of the ideal template is straightforward: the filtered
    sky map $\bm{D'}$ produced by the pipeline is
    \begin{align}\label{equ:filtered data}
        \bm{D'} = (\bm{I}-\bm{M}) \cdot \bm{F} \cdot (\bm{d}+\bm{d}_p),
    \end{align}
    where $\bm{d}_p$ is a single point source (assumed to have Gaussian
    shape in simulation) and
    \begin{align}
        \bm{d}=\bm{d}_c+\bm{d}_f+\bm{n}
    \end{align}
    is a column vector containing the input signals other than the point
    sources: $\bm{d}_c$ is the CMB signal, $\bm{d}_f$ is the foreground,
    and $\bm{n}$ is the noise. $\bm{F}$ is a square matrix representing
    the linear filtering effect. $\bm{M}$ is a diagonal matrix for the
    point source mask, which is 1 for the region around the point source
    and 0 elsewhere, and $\bm{I}$ is the identity matrix. Thus,
    $\bm{I}-\bm{M}$ is the non-point source region where the point
    source leakages need to be studied. For convenience, the filtered
    result of $\bm{d}$ is also computed as
    \begin{align}
        \bm{d'} = (\bm{I}-\bm{M}) \cdot \bm{F} \cdot \bm{d}.
    \end{align}
    Note that both $\bm{d}$ and $\bm{d}'$ are without the point source
    or point source leakages.

    It is clear from the descriptions above that the ideal template
    $\bm{\mathcal{T}}_0$ is
    \begin{align}\label{equ:ideal template}
        \bm{\mathcal{T}}_0 = (\bm{I}-\bm{M}) \cdot \bm{F} \cdot \bm{d}_{p},
    \end{align}
    which fully contains the point source leakage due to filtering
    except for an unknown point source amplitude\footnote{Strictly
    speaking, the point source polarization direction should also be
    taken into account, which will be managed in Sect.~\ref{sec:examples} by fitting the Q
    and U templates separately.}. If the amplitude of the template can be
    perfectly determined, then we have
    \begin{align}\label{equ:ideal template}
        \bm{D'} = \bm{d'} + \bm{\mathcal{T}}_0,
    \end{align}
    which separates the signal and point source leakages completely. In
    practice, the amplitude of the template should be determined by
    linear regression, and the best-fit template is subtracted to remove
    the point source leakage, leaving a residual that is no more than
    the chance correlation between $\bm{\mathcal{T}}_0$ and $\bm{d}'$,
    whose amplitude is typically a few percent of $\bm{d}'$.

    \subsection{The realistic template}\label{sub:realistic template}

    The ideal template can remove the point source leakages more
    effectively, but it necessitates precise knowledge of the beam
    profile, including the asymmetry, which is usually unavailable.
    Therefore, we go forward with creating a realistic template that
    can be obtained directly from the sky map produced by the pipeline.
    The main idea to construct the realistic template is based on three
    reasonable assumptions:
    \begin{enumerate}
        \item \label{itm:1}~~ The point sources are almost unaffected by
        the TOD filtering. This is true according to
        \cite{Ghosh:2022mje}, which shows the small scale structures are
        almost unaffected by the TOD filtering.

        \item \label{itm:2}~~The point source mask is big enough to
        include the majority of the point source. According to
        \cite{Li:2017drr, Salatino_2020}, the FWHM of AliCPT beam varies
        from $12'\sim19'$, which corresponds to the Gaussian beam width of $\sigma<10'$. Therefore,
        a mask of $r=40'$ region is enough to exclude most point
        sources, with an exception of a few extremely bright sources
        along the Galactic plane.

        \item \label{itm:3}~~The point source is significantly stronger
        than the CMB/foreground at the position of itself. Although it
        is possible to detect point sources that are weaker than the
        CMB, the leakages produced by these point sources are
        negligible, thus their leakages don't need to be taken into
        account.
    \end{enumerate}

    With assumption~\ref{itm:1}, it is easy to see that $\bm{d}_p
    \approx \bm{F} \cdot \bm{d}_{p}$, and assumption~\ref{itm:2} ensures
    $\bm{d}_p \approx
    \bm{M} \cdot  \bm{d}_{p}$; thus, we have:
    \begin{align}
        \bm{d}_p \approx \bm{M} \cdot \bm{F} \cdot \bm{M} \cdot  \bm{d}_{p}
    \end{align}
    Substitute the above one into Eq.~\ref{equ:ideal template}, we
    get the realistic template $\bm{\mathcal{T}}_1$ as:
    \begin{align}\label{equ:realistic template}
        \bm{\mathcal{T}}_1 = (\bm{I}-\bm{M}) \cdot \bm{F} \cdot (\bm{M}
        \cdot \bm{F} \cdot \bm{M} \cdot \bm{d}_{p})\approx
        \bm{\mathcal{T}}_0
    \end{align}   
    
    The above equation is crucial, because $(\bm{M} \cdot \bm{F} \cdot
    \bm{M} \cdot \bm{d}_{p})$ is the sky map produced by the pipeline
    with a point source mask $\bm{M}$\footnote{Assumption~\ref{itm:3}
    ensures that the CMB is negligible compared with the point sources
    at their positions.}. Thus, no need for us to know the details of the beam profile, and $\bm{\mathcal{T}}_1$ can be obtained by
    feeding the sky map product of the pipeline (with a proper point
    source mask applied) into the pipeline again as $\bm{F} \cdot
    (\bm{M}\cdot \bm{F} \cdot \bm{M} \cdot \bm{d}_{p})$.

    \subsection{Other procedures}\label{sub:other procs}
    
    The aforementioned method is firstly tested using a single point
    source simulation. First, Eq.~\ref{equ:filtered data} constructs the filtered sky map, and Eqs. \ref{equ:ideal template}\&~\ref{equ:realistic template} provide the ideal and realistic
    templates $\bm{\mathcal{T}}_0$ and $\bm{\mathcal{T}}_1$,
    respectively. If the polarization direction of a point source is
    already known, since the leakage is caused by both of Q and U in the
    pipeline, a fitting parameter $k$ for both of Q and U is determined
    using linear regression to minimize the RMS (root-mean-square) of
    residual. Here the RMS of residual is:
    \begin{align}\label{equ: sigma basic}
        (\sigma^Q_\xi)^2 &= \frac{1}{n} \left[ \sum_{i=1}^n (\bm{D'} - k_\xi \cdot \bm{\mathcal{T}}_\xi)^2_i \right]_Q\,, \\ \nonumber
        (\sigma^U_\xi)^2 &= \frac{1}{n} \left[ \sum_{i=1}^n (\bm{D'} - k_\xi \cdot \bm{\mathcal{T}}_\xi)^2_i \right]_U\,, \\ \nonumber
        \sigma^2_\xi &= (\sigma^Q_\xi)^2 + (\sigma^U_\xi)^2
    \end{align}
    where $i$ is pixel number, $n$ is the total number of pixels, the superscripts $Q$ and $U$ of $\sigma$ denote the Stokes parameters, the subscript $Q$ or $U$ on the right term stands for the Stokes parameter vector solely considered in this formula, and $\xi =0, 1$ for the ideal or realistic templates respectively.
    
    If the point source's polarization direction is unknown, we
    construct separately $\bm{\mathcal{T}}^Q$ and $\bm{\mathcal{T}}^U$,
    whose input only contains values for $Q$ (or $U$) Stokes parameter
    vector. Multi-linear regression should be applied to simultaneously
    determine $k^Q$ and $k^U$ in order to minimize the RMS of residual
    in the pixel domain. This is because the filtering calculation will
    mix the Q and U Stokes parameter, and $\bm{\mathcal{T}}^Q$ (or
    $\bm{\mathcal{T}}^U$) is not zero on $U$-part (or $Q$-part) even 
    for a $Q$-only (or $U$-only) input. Hence for each of the Stokes 
    parameters, the RMS of the residual is as follows:
    \begin{align}\label{equ: sigma basic unknown}
        (\sigma_{\xi}^Q)^2 &= \frac{1}{n} \left[ \sum_{i=1}^n (\bm{D'} - k_{\xi}^Q\cdot \bm{\mathcal{T}}_{\xi}^Q - k_{\xi}^U\cdot \bm{\mathcal{T}}_{\xi}^U)^2_i \right]_Q\,, \\ \nonumber
        (\sigma_{\xi}^U)^2 &= \frac{1}{n} \left[ \sum_{i=1}^n (\bm{D'} - k_{\xi}^Q\cdot \bm{\mathcal{T}}_{\xi}^Q - k_{\xi}^U\cdot \bm{\mathcal{T}}_{\xi}^U)^2_i \right]_U\,, \\ \nonumber
        \sigma_{\xi}^2 &= (\sigma_{\xi}^Q)^2 + (\sigma_{\xi}^U)^2
    \end{align}
    where the Stokes parameter vector is solely taken into account in
    the formula by the subscript $Q$ (or $U$) of the right term.
    
    We also test our method with multiple point sources. The filtered
    sky map $\bm{D'}$ is shown here as follows:
     \begin{align}\label{equ:filtered data multi}
        \bm{D'} = (\bm{I}-\bm{M}_{\Sigma}) \cdot \bm{F} \cdot (\bm{M}_{\Sigma} \cdot \sum_{j} \bm{d}_{p,j}+\bm{d})\,,
     \end{align}
    where $j$ is the index of point source, $\bm{M}_{\Sigma} =
    \sum_{j}\bm{M}_{j}$ is the mask for all point
    sources\footnote{Assume the point sources are non-overlapping,
    otherwise the summation should be replaced by the "exclusive or" (XOR) operation.}
    and $\bm{M}_{j}$ is the $j^{th}$ single point source mask, and
    $\bm{I}-\bm{M}_{\Sigma}$ is the non-point source region to
    investigate the impact of leakage. Hence the ideal and realistic
    templates  for each point source are:
    \begin{align}
       \bm{\mathcal{T}}_{0,j} &= (\bm{I}-\bm{M}_{\Sigma}) \cdot \bm{F} \cdot \bm{M}_{j} \cdot \bm{d}_{p,j}\,, \\ \nonumber
       \bm{\mathcal{T}}_{1,j} &= (\bm{I}-\bm{M}_{\Sigma}) \cdot \bm{F} \cdot (\bm{M}_{j} \cdot \bm{F} \cdot \bm{M}_{j} \cdot  \bm{d}_{p,j})
    \end{align}
    Correspondingly, the fitting parameters $k_{\xi j}$ (when the
    polarization direction is known) or $k_{\xi j}^Q$ and $k_{\xi j}^U$
    (when the polarization direction is unknown) are computed for each
    point source using a multi-linear regression that takes into account
    all the templates simultaneously, in order to minimize the RMS of
    residuals in the combined non-point sources region. If the
    polarization directions of point sources are already known, then the
    RMS are:
    \begin{align}
       (\sigma_\xi^Q)^2 &= \frac{1}{n} \left[ \sum_{i=1}^n (\bm{D'} - \sum_{j} k_{\xi j} \bm{\mathcal{T}}_{\xi j})^2_i\right]_Q\,, \\ \nonumber
       (\sigma_\xi^U)^2 &= \frac{1}{n} \left[ \sum_{i=1}^n (\bm{D'} - \sum_{j} k_{\xi j} \bm{\mathcal{T}}_{\xi j})^2_i\right]_U \,,\\ \nonumber
       \sigma^2_\xi &= (\sigma_\xi^Q)^2 + (\sigma_\xi^U)^2
    \end{align}
    and if the polarization directions are unknown, then the RMS
    are:
    \begin{align}
       (\sigma_\xi^Q)^2 &= \frac{1}{n} \left[ \sum_{i=1}^n (\bm{D'} - \sum_{j} k_{\xi j}^Q \bm{\mathcal{T}}_{\xi j}^Q - \sum_{j} k_{\xi j}^U \bm{\mathcal{T}}_{\xi j}^U)^2_i\right]_Q \,,\\ \nonumber
       (\sigma_\xi^U)^2 &= \frac{1}{n} \left[ \sum_{i=1}^n (\bm{D'} - \sum_{j} k_{\xi j}^Q \bm{\mathcal{T}}_{\xi j}^Q - \sum_{j} k_{\xi j}^U \bm{\mathcal{T}}_{\xi j}^U)^2_i\right]_U \,,\\ \nonumber
       \sigma^2_\xi &= (\sigma_\xi^Q)^2 + (\sigma_\xi^U)^2 
    \end{align}
  
\section{Simulations and tests}\label{sec:examples}
  In the computation that follows, we use the outcome with the
   local monopoles subtracted from each Stokes parameter in the
   non-point source region, in order to demonstrate the method's ability
   to correct the leakages. To reduce accidental fluctuation, we also use
   10 different CMB and noise realizations in the simulation.
   
   After determining the fitting parameters by multi-linear regression,
   we build dB ($20\log_{10}P$) sky maps of the two templates, their
   residuals and compute the dB effect to compare the degree of
   correction in the pixel domain. All these dB maps are computed from
   the polarization intensity $P$.

   The residual map is the difference between the filtered sky map (no
   leakage effect from the beginning) and cleaned sky map. When the
   polarization directions are already known, the residual map is:
   \begin{align} \label{eq:res}
     \bm{\delta}_{\xi} = \left(\bm{D'} - \sum_{j} k_{\xi j} \bm{\mathcal{T}}_{\xi j}\right) - \bm{d'}\,,
   \end{align}   
   and when the polarization direction is unknown, the residual is:
   \begin{align} \label{eq:res1}
     \bm{\delta}_{\xi} = \left(\bm{D'} - \sum_{j} k^Q_{\xi j} \bm{\mathcal{T}}^Q_{\xi j}  - \sum_{j} k^U_{\xi j} \bm{\mathcal{T}}^U_{\xi j}\right) - \bm{d'}\,,
   \end{align}   
   where $\xi =0, 1$ for the ideal and realistic templates, respectively. Now we introduce a quantity, $\bm{E}^{\rm dB}_{\xi ,P}$, dubbed as "dB effect", for easily comparing the residuals with respect to the CMB in units of $\rm dB$, in the form of:
   \begin{align}\label{eq:db}
   \bm{E}^{\rm dB}_{\xi ,P} = 20\log_{10} \dfrac{P_{\bm{\delta}}}{\langle P_{\rm cmb} \rangle}\,,
   \end{align}   
   where the superscript $\rm dB$ stands for the units of decibels, and $P$ is the polarization intensity defined as $P = \sqrt{Q^2 + U^2}$. $\langle P_{\rm cmb} \rangle$ is the mean value of CMB polarization intensity, about $2.070$ $\mu$K.

   \subsection{Single point source}
      For a single point source, the fitting parameter for the ideal
      template is very close to 1, whereas the fitting parameter for
      realistic template is above 1, because the amplitude of the point
      source is reduced by filtering. The fitting residuals are 1 to 2
      orders of magnitudes less than the point source leakage template.
      
      We select a location at $[b,l] = [50^{\circ}, 190^{\circ}]$,
      assuming that its polarization intensity is equal to $150$ $\mu
      K$. We then smooth this point by $19'$ to make a Gaussian point
      source. We first assume that the polarization direction is already
      known, where we fix the polarization angle of this point source to
      be $22.5^{\circ}$, resulting in a Gaussian point source with the
      Q, U values of approximately $106$ $\mu$K. For this artificially
      point source, the fitting parameter with true leakage is close to
      0.997 and polarization residual standard deviation is approximately 
      $1.113\times 10^{-3}$ $\mu$K in the pixel domain. The fitting 
      parameter with realistic template is close to 1.201, and the 
      polarization residual standard deviation is roughly
      $1.999\times 10^{-3}$ $\mu$K in the pixel domain. Meanwhile, the true 
      leakage and the realistic template have standard deviations of $0.028~\mu$K and $0.024~\mu$K for polarization, respectively. Additionally, after filtering 
      calculation, polarization standard deviations of the CMB, foreground 
      and noise are, in that order, $0.374$ $\mu$K, $0.027$ $\mu$K, 
      and $0.121$ $\mu$K. After using our method for alleviating the 
      impact of point source leakage, the standard deviation of residuals is consequently orders of magnitudes smaller than that of template 
      we build, as well as smaller than that of CMB, foreground and noise. 
      The impact of point source leakage is lessened by our correction method.
      
\begin{figure}[htbp]
        \centering
        \includegraphics[width=0.3\textwidth]{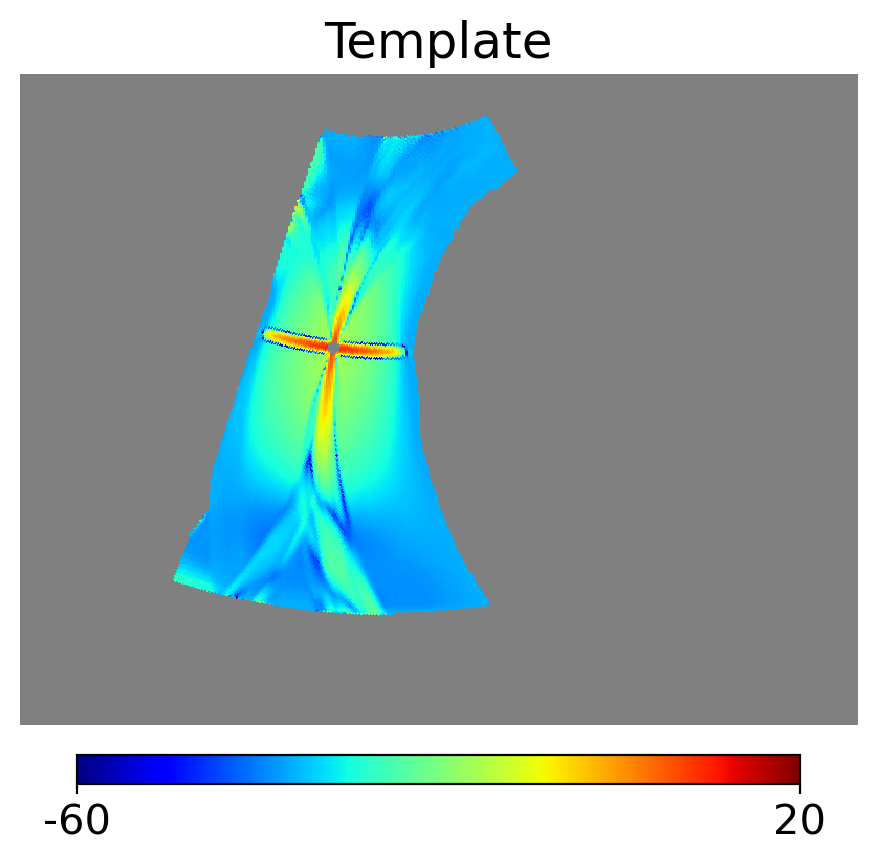}
        \includegraphics[width=0.3\textwidth]{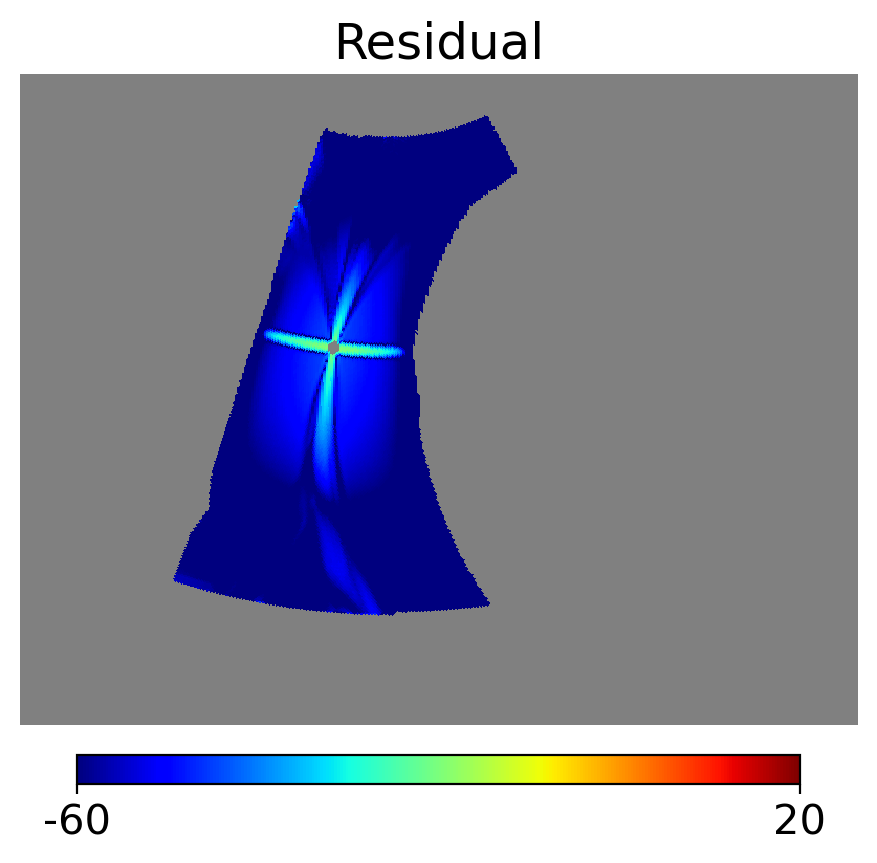}
        \includegraphics[width=0.3\textwidth]{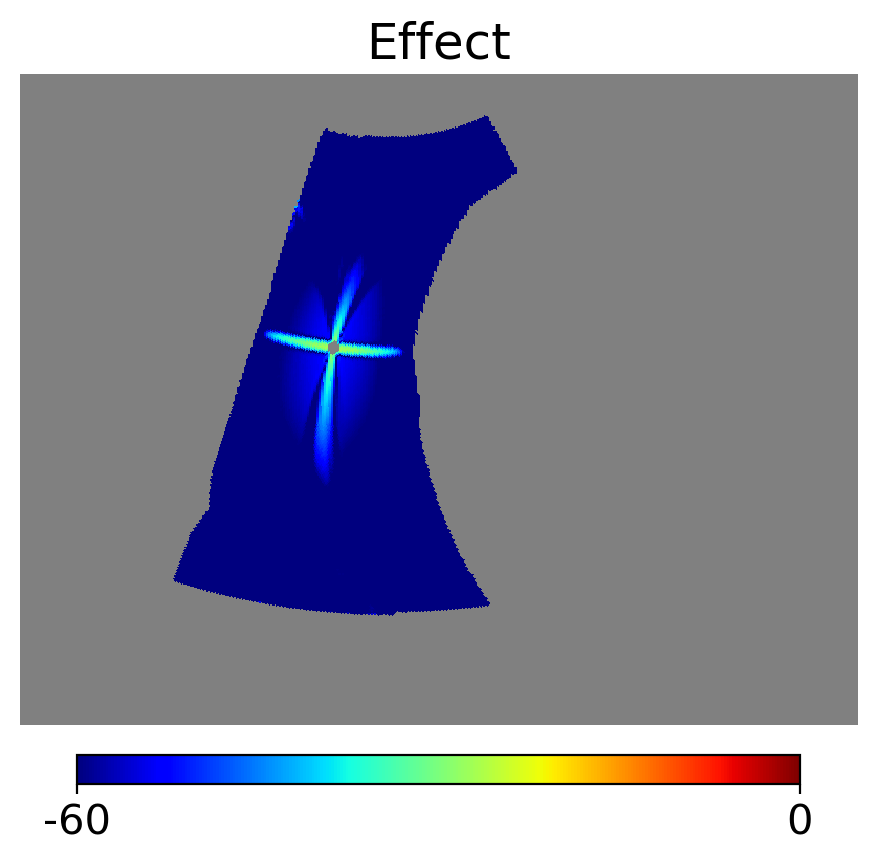}
        \includegraphics[width=0.3\textwidth]{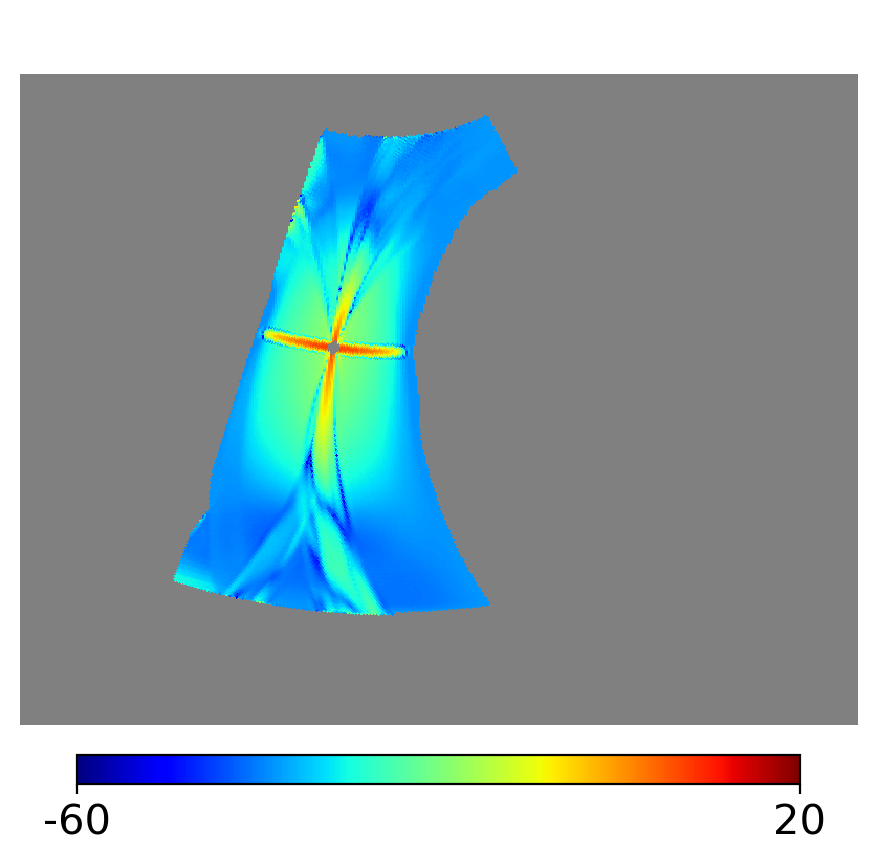}
        \includegraphics[width=0.3\textwidth]{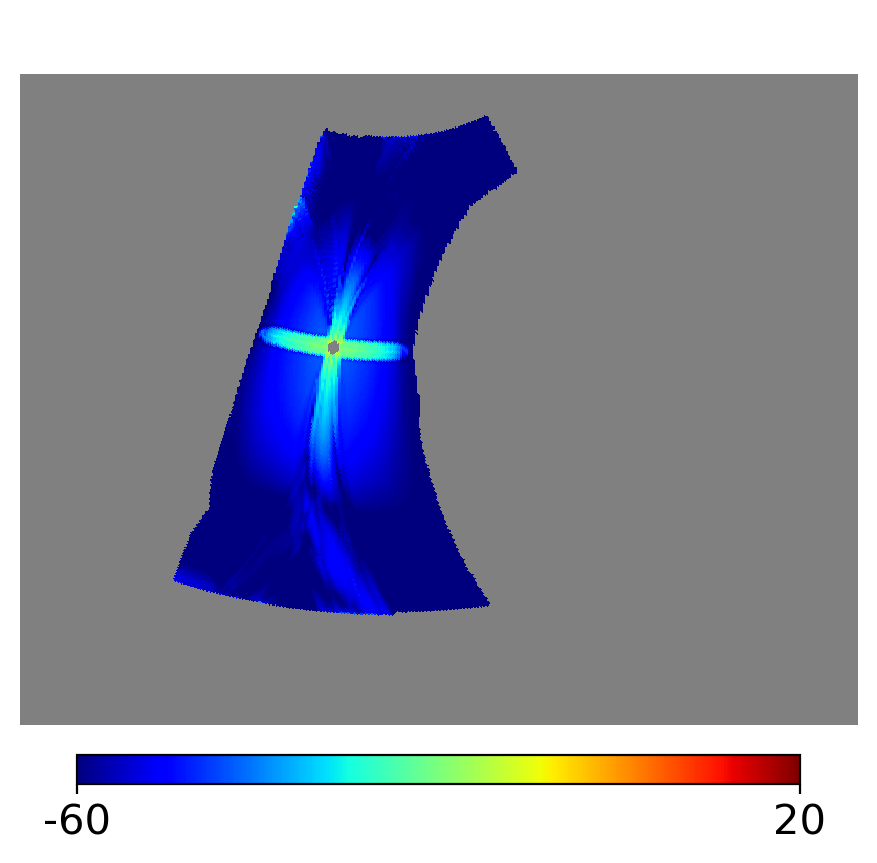}
        \includegraphics[width=0.3\textwidth]{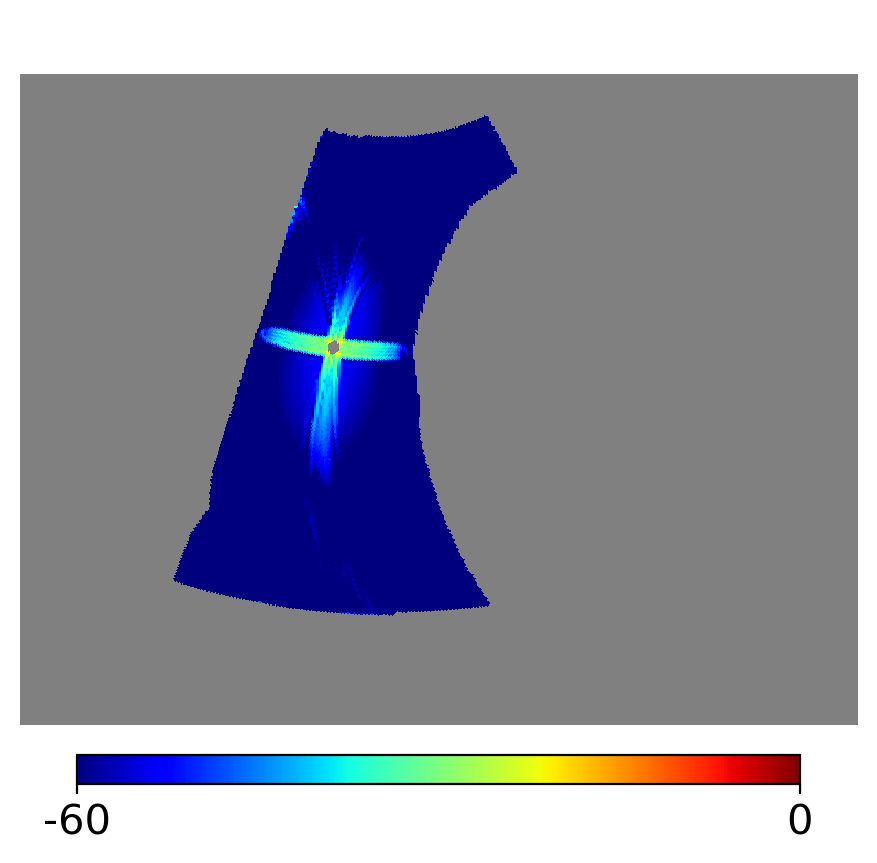}
        \caption{Comparison of results in units of decibels (dB), obtained using the ideal template (upper) and the realistic template (lower), for the case of a single point source with a given polarization direction, where the polarization values of the templates (left), the residuals (middle) from Eq.~\ref{eq:res} and the dB effects (right) as defined in Eq.~\ref{eq:db} are shown, respectively.}
        \label{fig:single sky map}
      \end{figure}

      For a single Gaussian point source with a given polarization
      direction, we now present the polarization dB value of templates,
      residuals and dB effects, as illustrated in Fig. \ref{fig:single
      sky map}. The leakage of point source is diffused on the partial
      sky map after pipeline, as seen in the right column of Fig.
      \ref{fig:single sky map}, and it resembles a star. Although the
      residual's amplitude is frequently lower than that of template,
      the residual also exhibits vivid spikes on the sky map (middle
      column). Additionally, dB effect shows the proportion between
      residual and mean CMB polarization. Realistic dB effect error is
      more than that of ideal dB effect (right column). Our method
      provides good accuracy in the case of single Gaussian point source
      because the residual left over after correction procedure is
      significantly less than the true leakage.
      
\begin{figure}[htbp]
        \centering
        \includegraphics[width=0.45\textwidth]{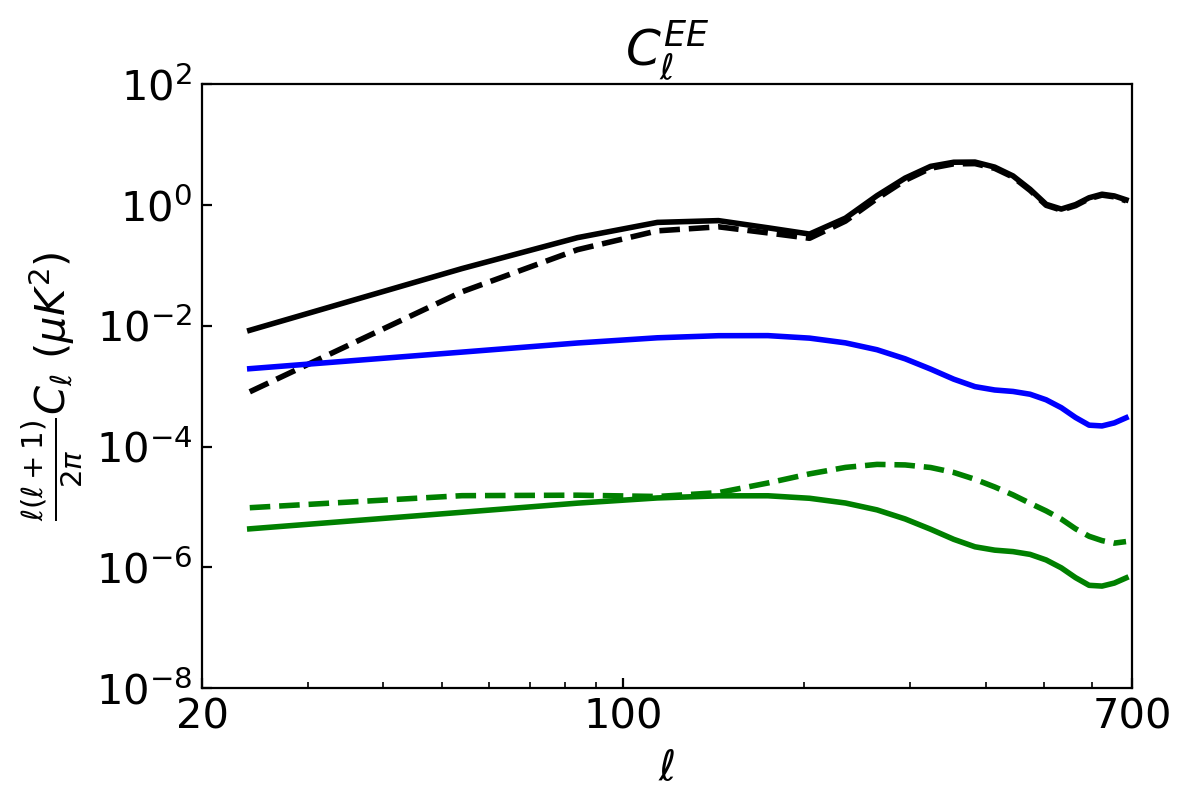}
        \includegraphics[width=0.45\textwidth]{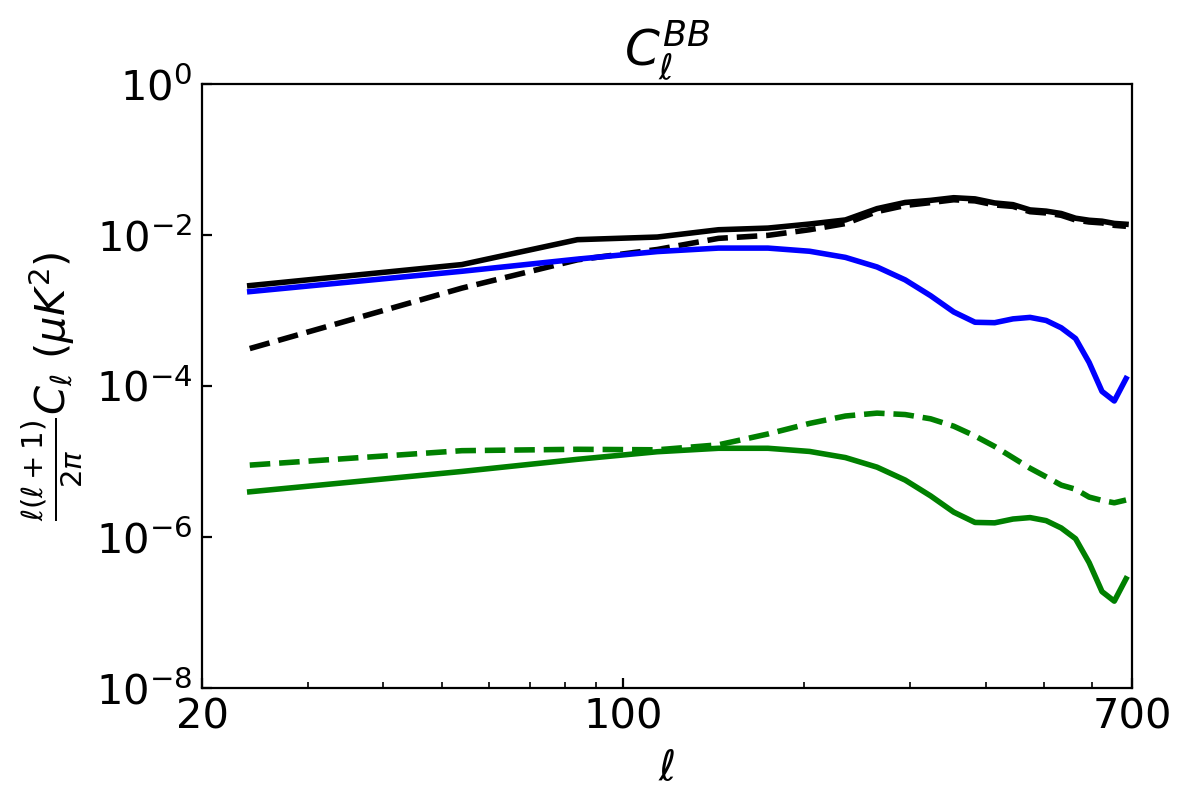}
        \caption{$EE$ and $BB$ power spectra for single point source with given polarization direction, including input (black solid) and filtered CMB (black dashed), residuals after the leakage removal for the ideal template (green solid) and the realistic template (green dashed), and the true leakage (blue).}
        \label{fig:single cl}
      \end{figure}

      For single point source with known polarization direction, we compute the angular power spectrum of CMB, residual, and true leakage, as shown in Fig. \ref{fig:single cl}. We select $\ell$ with a range of 20 to 700, because we only focus on part of the sky map. The residual spectra are 2 or 3 orders of magnitudes lower than that of the templates. For the $BB$ spectrum, the spectrum of residual is significantly smaller after our correction method than that of CMB. Hence our method can help to reduce the measurement error.
      
      The impact of unknown polarization direction is then verified in the case of single point source. On the basis that the polarization direction of this point source was known, the fitting parameter was previously determined. Assuming the point source's polarization direction is unknown, the parameters are to be determined by fitting. We continue to construct the same ideal and realistic template as Eqs.~\ref{equ:ideal template}\&~\ref{equ:realistic template}. For each type of template, $k^U/k^Q$ is equal to $\tan(2\theta)$ because that the filtering procedure is equivalent to linear matrix multiplication calculation. Taking $\theta = 33.9^\circ$ as an example, the result of fitting parameters, residual standard deviations and the comparison of ratios under the condition of unknown polarization direction are presented in Tab. \ref{tab:single polarized}. For both ideal and realistic templates, $k^U/k^Q$ is nearly equal to $\tan(2\theta)$. The accuracy is affected to some extent by the CMB, foreground and noise due to the chance correlation. It demonstrates fitting parameters are still stable for the point source whose polarization direction is unknown.
      
      \begin{table}[!h]
         \caption{ Derived fitting parameters for single point source with unknown polarization direction.}
         \centering
         \begin{tabular}{ccccccccc}
            \toprule
             $k_0^Q$ & $k_0^U$ & $\sigma_{\delta_0} /10^{-3} \mu$K & $k_1^Q$ & $k_1^U$ & $\sigma_{\delta_1} /10^{-3} \mu$K & $k_0^U/k_0^Q$ & $k_1^U/k_1^Q$ & $\tan(2\theta)$\\
            \midrule
             0.534 & 1.303 & 2.047 & 0.645 & 1.567 & 2.600 & 2.439 & 2.429 & 2.453 \\
            \bottomrule
         \end{tabular}
         \label{tab:single polarized}
      \end{table}

      As a result of filtering calculation, there is a star-like leakage for a single Gaussian point source. Additionally, our approach can lower pixel domain error and make the residual spectrum much lower than the CMB spectrum. Furthermore, our method is still reliable and accurate even when the polarization direction of point source is unknown.
          
   \subsection{Multiple point sources}
      In the case of multiple point sources, the multi-linear regression method's fitting parameters for each point source are comparable to the fitting results when each point source is treated as a single point source. The fitting residuals are still 1 to 2 orders of magnitudes less than the point source's true leakages.
      
      The same approach as the case of single point source is used on five point sources with random locations and smooth with $19'$. Tab.~\ref{tab:param compare} displays their locations and polarization amplitudes. For each point source, we construct accordingly the ideal and realistic templates. Under the assumption of given polarization direction, the fitting parameters for multiple point sources are given in Tab. \ref{tab:param compare}, and are almost identical to the fitting parameters for each single point source, demonstrating the stability of our correction method for multiple point sources.
      \begin{table}[!h]
         \caption{Derived fitting parameters for multiple point sources by the multi-linear regression method when the polarization direction is known for each point source, which are comparable with those for each single point source.}
         \centering
         \begin{tabular}{ccccc}
            \toprule
            Number & $[b, l]^{\circ}$ & $P/\mu$K & $k_0$ & $k_1$ \\
            \midrule
            1 & [50, 190] & 150 & 0.997 & 1.201 \\
            2 & [27, 191] & 200 & 0.995 & 1.187 \\
            3 & [30, 173] & 150 & 1.105 & 1.189 \\
            4 & [39, 162] & 120 & 1.021 & 1.199 \\
            5 & [56, 160] &  90 & 1.021 & 1.282 \\
            \bottomrule
         \end{tabular}
         \label{tab:param compare}
      \end{table}
      
We also examine the influence of polarization direction of multiple point sources. In order to simulate a filtered sky map $\bm{D'}$ for multiple point sources with unknown polarization directions, a series of polarizing angles $\theta_{j}$ are applied to the Stokes parameters of each individual point source $\bm{d}_{p,j}$ in Eq.~\ref{equ:filtered data multi}, where $j$ is the index of each point source. The result of fitting parameters for multiple point sources with unknown polarization direction is provided in Tab. \ref{tab:multi polarized}. The result is consistent with that for single point source. For each point source, the absolute values of $k^Q_{1}$ and $k^U_{1}$ are slightly higher than those of $k^Q_{0}$ and $k^U_{0}$ respectively, but $k^U/k^Q$ is still almost equal to $\tan(2\theta)$ for both the ideal and realistic templates, and the CMB, foreground, and noise still have a minor impact on the residual. Thus, for multiple point sources with unknown polarization direction, our method is still robust.
    \begin{table}[!h]
       \caption{Derived fitting parameters for multiple point sources with unknown polarization direction.}
       \centering
       \begin{tabular}{ccccccccc}
          \toprule
          Number & $\theta^{\circ}$ & $k_0^Q$ & $k_0^U$ & $k_1^Q$ & $k_1^U$ & $k_0^U/k_0^Q$ & $k_1^U/k_1^Q$ & $\tan(2\theta)$ \\
          \midrule
          1 & 33.9 &  0.535 & 1.304 &  0.646 & 1.568 &  2.438 &  2.428 &  2.453 \\
          2 & 10.8 &  1.311 & 0.513 &  1.564 & 0.613 &  0.392 &  0.392 &  0.395 \\
          3 & 19.3 &  1.164 & 0.854 &  1.363 & 1.000 &  0.734 &  0.734 &  0.796 \\
          4 & 56.4 & -0.580 & 1.377 & -0.682 & 1.616 & -2.373 & -2.368 & -2.374 \\
          5 & 24.7 &  1.015 & 1.023 &  1.273 & 1.285 &  1.008 &  1.010 &  1.165 \\          
          \bottomrule
       \end{tabular}
       \label{tab:multi polarized}
    \end{table} 
    
\begin{figure}[htbp]
        \centering
        \includegraphics[width=0.3\textwidth]{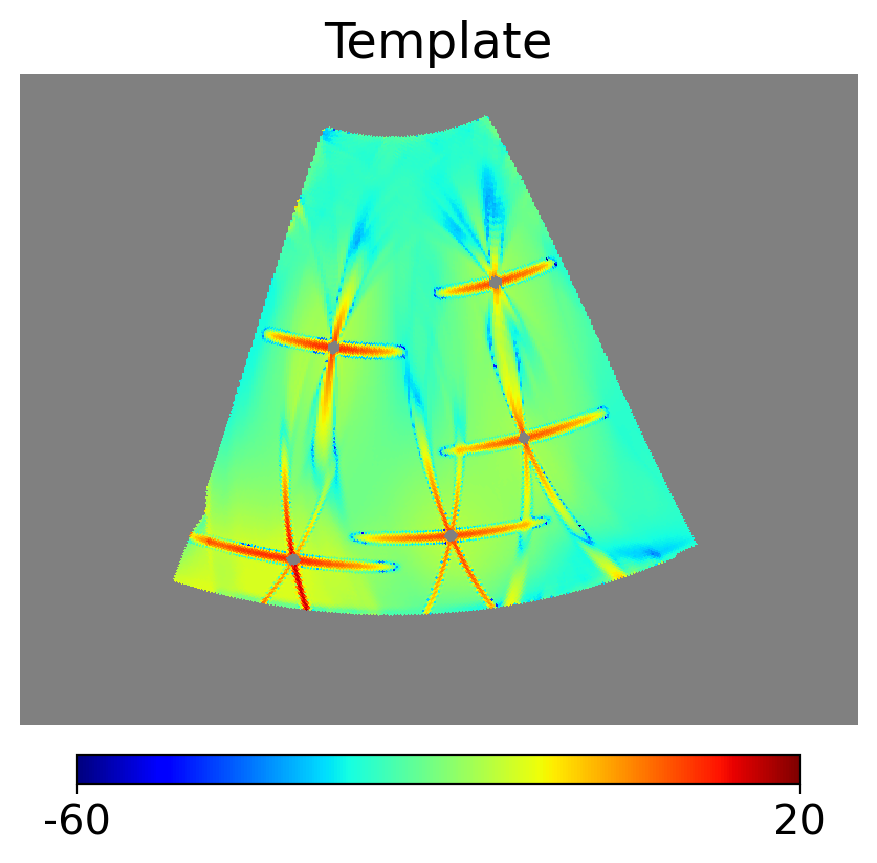}
        \includegraphics[width=0.3\textwidth]{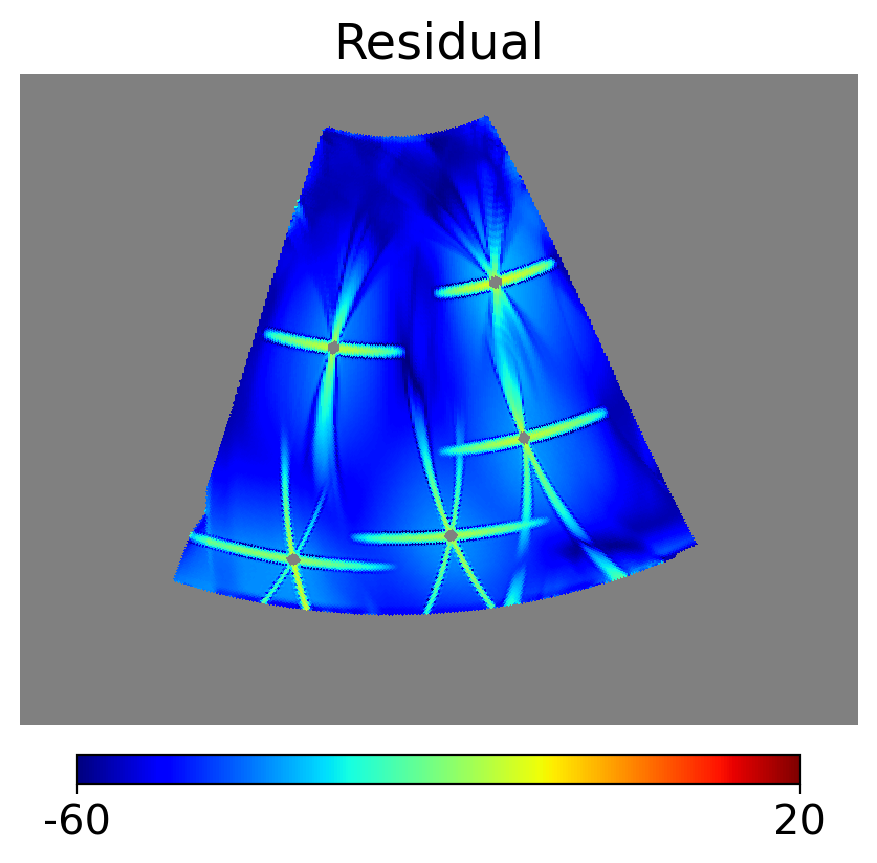}
        \includegraphics[width=0.3\textwidth]{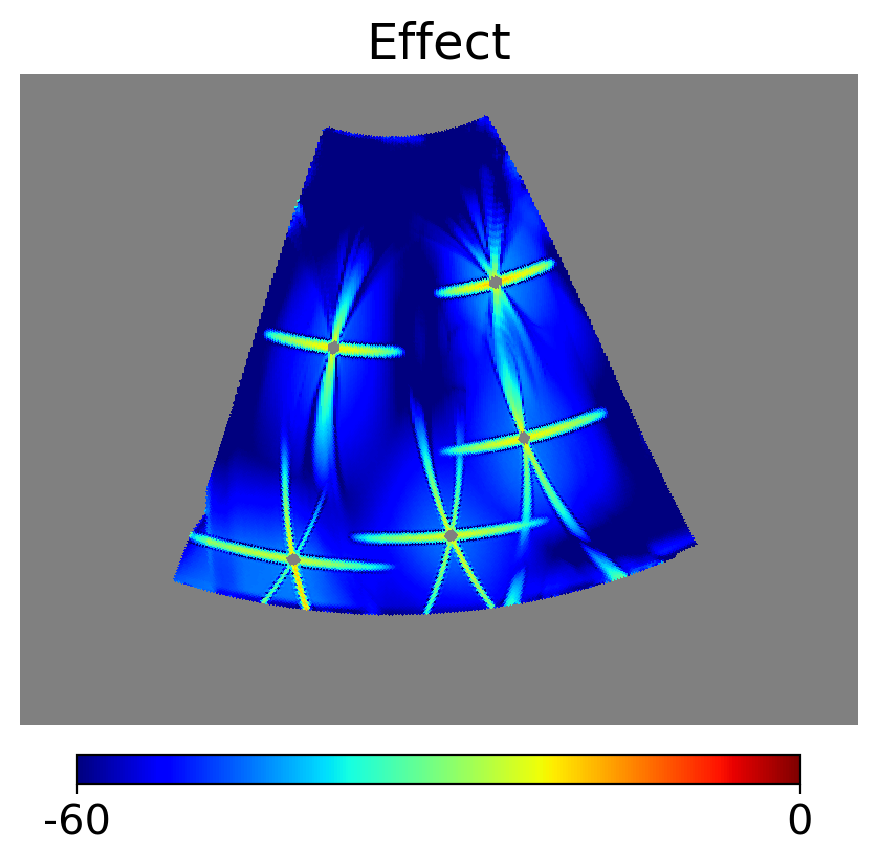}
        \includegraphics[width=0.3\textwidth]{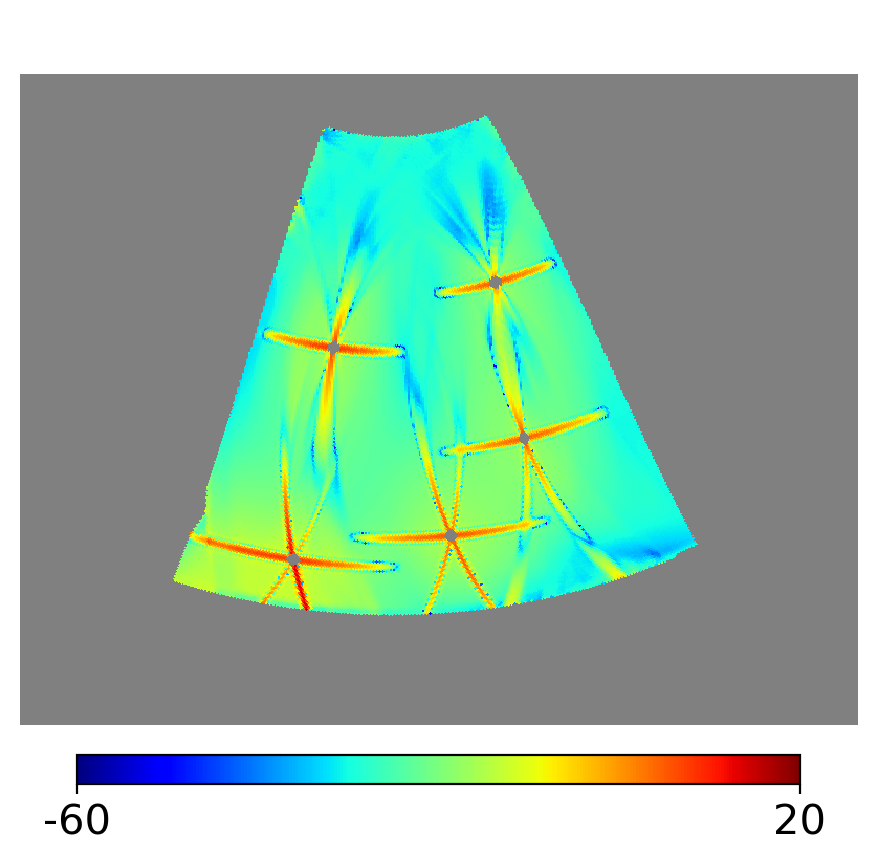}
        \includegraphics[width=0.3\textwidth]{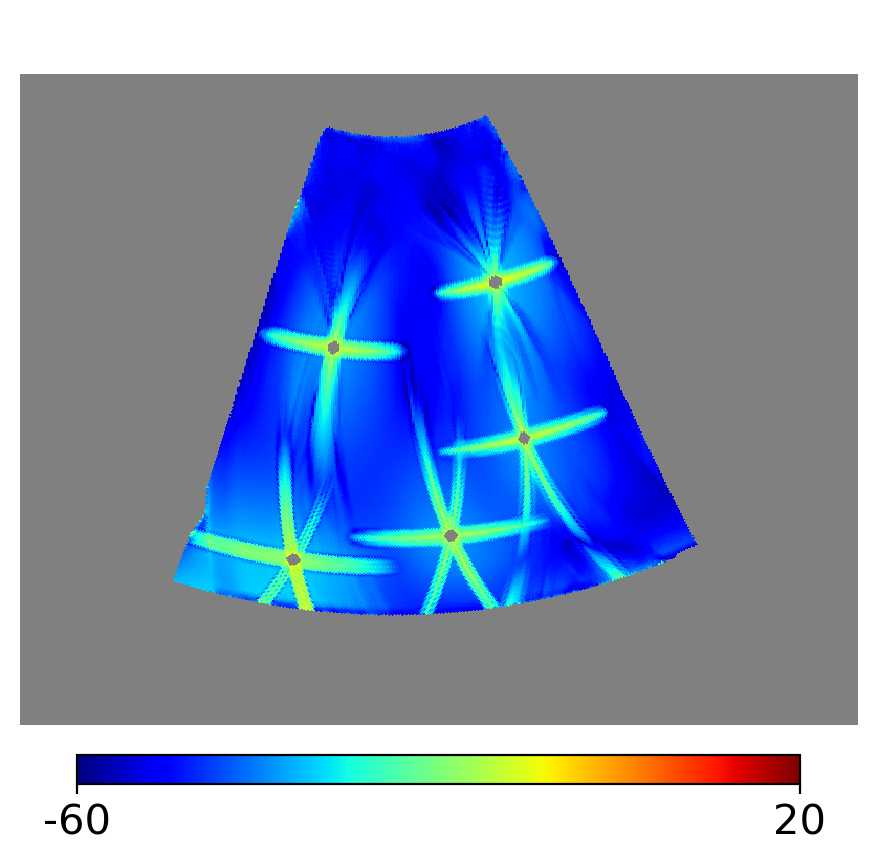}
        \includegraphics[width=0.3\textwidth]{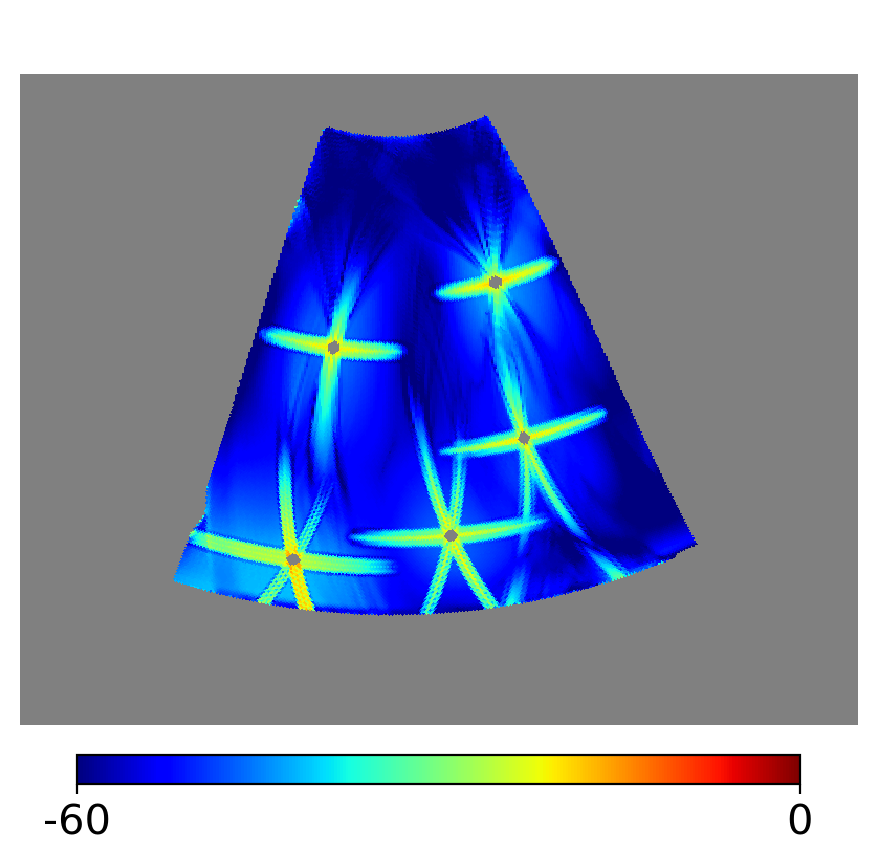}
        \caption{Same as in Fig.~\ref{fig:single sky map}, but for the case of multiple point sources with unknown polarization direction, where the residuals
        are estimated through Eq.~\ref{eq:res1}.} 
        \label{fig:multi sky map}
      \end{figure}

      We again compare the true leakage, realistic template, residuals, and dB effects for multiple Gaussian point sources with unknown polarization directions in Fig. \ref{fig:multi sky map}. Each point source has a leakage that resembles a star after the pipeline, and these leakages have an impact on the entire non-point source region we study. The middle column shows that residuals after the correction still exhibit star-like pattern surrounding the point sources, but with a substantially smaller amplitude. Furthermore, dB effect shows the relative effect of correction.
      
\begin{figure}[htbp]
        \centering
        \includegraphics[width=0.45\textwidth]{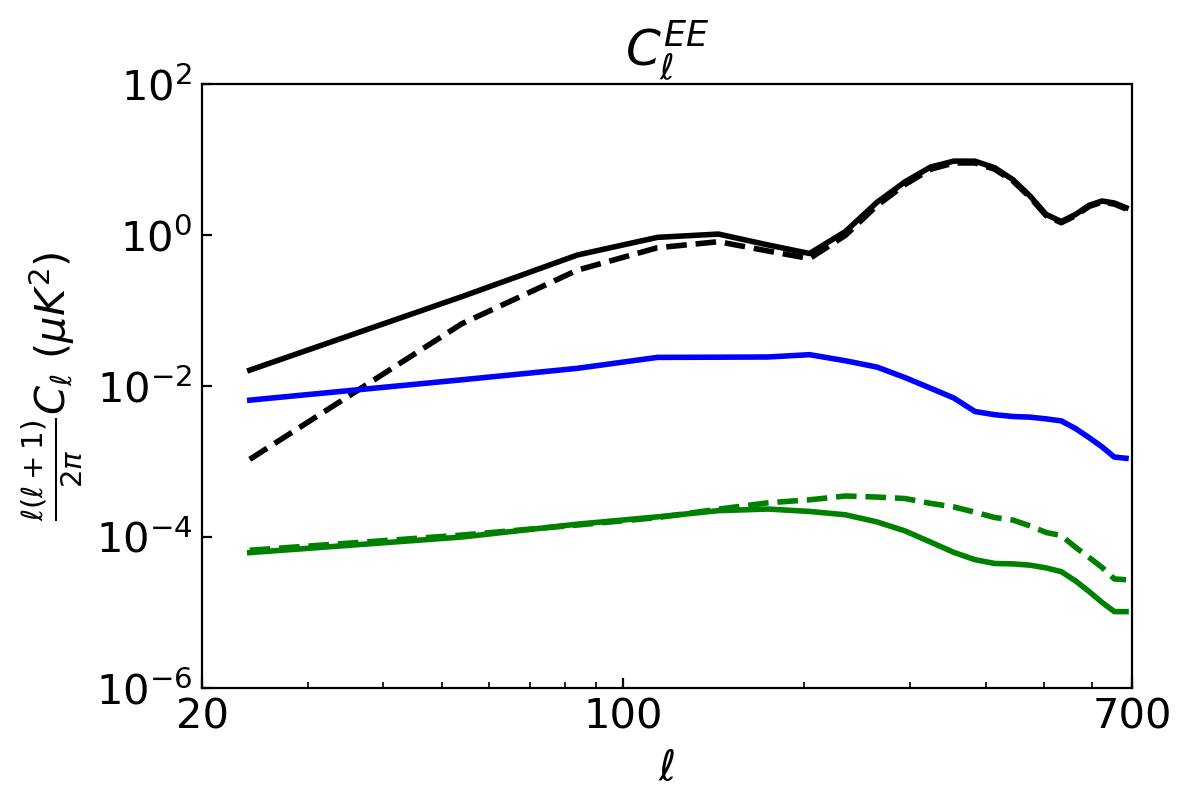}
        \includegraphics[width=0.45\textwidth]{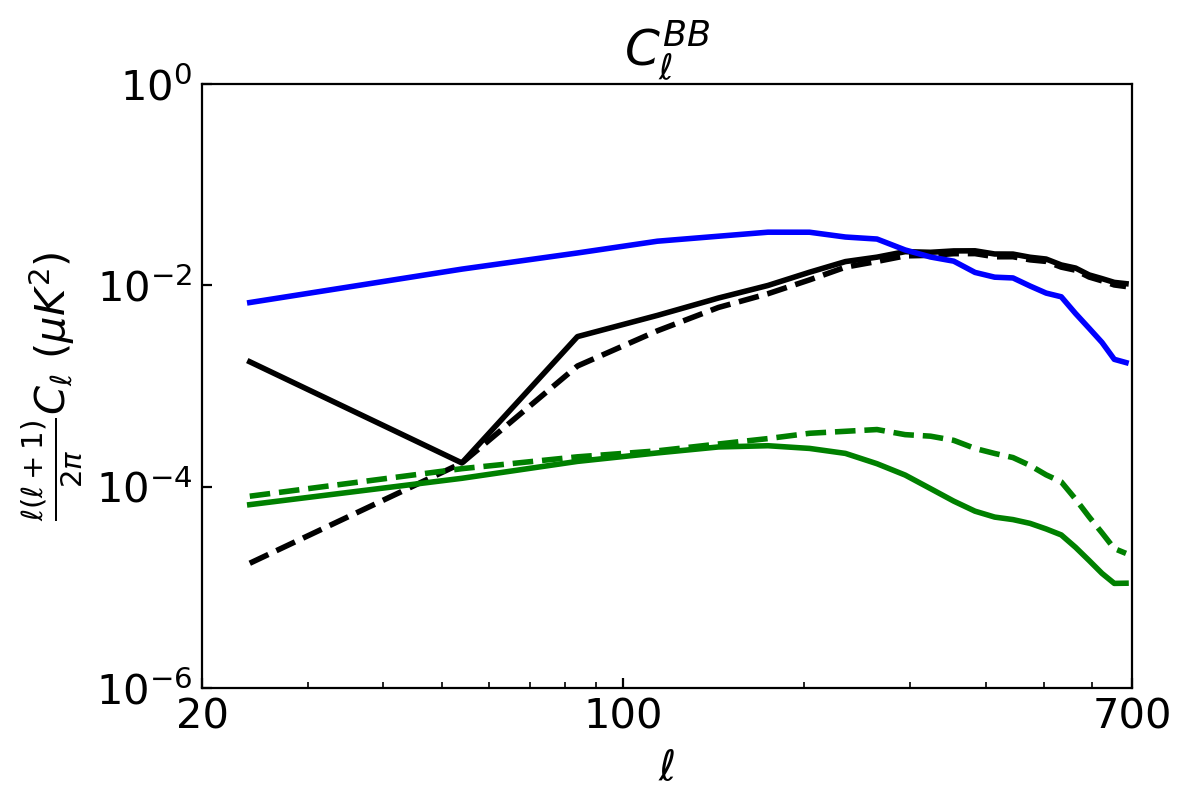}
        \caption{Same as in Fig.~\ref{fig:single cl}, but for
        the case of multiple point sources with unknown polarization direction.}
        \label{fig:multi cl}
      \end{figure}

      As shown in Fig. \ref{fig:multi cl}, using a logarithmic scale, we also compute the angular power spectrum of CMB, residuals, and true leakage for multiple point sources with unknown polarization directions. When compared to the spectrum of two templates, the spectrum of two residuals is 2 to 3 orders of magnitudes smaller. Our method can improve the performance of the probe since the $BB$ spectrum of residual is substantially smaller after our correction method than that of CMB. In addition, there are some fluctuations of the $BB$ power spectrum for CMB in small multipole $\ell$ as we only study a portion of sky map. Furthermore, the power spectrum with unknown polarization direction is also a little greater than that with known polarization direction. This is because two fitting parameters can introduce more errors, which causes the residual to increase.
    
      In conclusion, even when the number of point sources rises, our correction method maintains good accuracy and reliability. Regardless of whether there are one or more point sources, fitting parameters are almost identical. The amplitude of residuals after correction is much lower than that of the template in the pixel domain. After correction, the residual spectrum is considerably smaller than the CMB spectrum, thus the impact of point source leakage is significantly reduced by our method.
      
   \subsection{Actual point sources}
    To simulate the actual sky map\footnote{https://pla.esac.esa.int/\#catalogues.}, we use the data from 2013 Planck Catalogue of Compact Sources(PCCS). In the region we study (the same region as in Fig. \ref{fig:multi sky map}), the flux data of  the 10 brightest point sources at 100 GHz are used to calculate the conversion coefficient from flux to temperature, which is equal to 2.879 $\mu$K$\cdot\mathrm{mJy}^{-1}$\footnote{Due to the different point source spectra, the standard HFI unit conversion coefficient (4.583 $\mu$K$\cdot\mathrm{mJy}^{-1}$) in the Planck 2013 result~\citep{2014A&A...571A...9P} is slightly different from the value used here.}. The approximate temperature of these 10 brightest point sources is then determined. The point source polarization is assumed to be roughly 40 percent of its temperature. Since the precise point source polarization directions are unknown before we get the corrected map, a certain polarization direction is specified for each point source to construct two types of templates for simplification of calculation. According to the fitting parameter results in Tab. \ref{tab:actual polarized}, for both the ideal and realistic templates, the ratio of fitting parameters $k^U/k^Q$ for each point source is close to $\tan(2\theta)$.
    \begin{table}[!h]
         \caption{Location, temperature, polarization and fitting parameter result with random polarization direction of actual 10 brightest point sources in the region we study (the region shape is the same with Fig. \ref{fig:multi sky map}).}
         \centering
         {\tiny
         \begin{tabular}{ccrrrrrrrrrr}
            \toprule
            Number & $[b, l]^{\circ}$ & $I/\mu$K & $P/\mu$K & $\theta^{\circ}$ & $k_0^Q$ & $k_1^Q$ & $k_0^U$ & $k_1^U$ & $k_0^U/k_0^Q$ & $k_1^U/k_1^Q$ & $\tan(2\theta)$ \\
            \midrule
            1  & [46.2, 183.7] & 880.4 & 352.1 & 155.8 &  0.939 & -1.039 &  1.119 & -1.237 & -1.107 & -1.106 & -1.125 \\
            2  & [31.9, 200.0] & 378.2 & 151.3 & 25.0  &  0.927 &  1.074 &  1.098 &  1.271 &  1.159 &  1.157 &  1.191 \\
            3  & [44.8, 175.7] & 331.0 & 132.4 & 25.8  &  0.914 &  1.100 &  1.084 &  1.306 &  1.203 &  1.205 &  1.258 \\
            4  & [58.5, 177.6] & 232.3 &  92.9 & 20.1  &  1.124 &  0.785 &  1.433 &  0.999 &  0.698 &  0.697 &  0.844 \\
            5  & [22.8, 196.5] & 186.0 &  74.4 & 153.3 &  0.774 & -1.096 &  1.042 & -1.516 & -1.416 & -1.455 & -1.351 \\
            6  & [33.3, 178.2] & 185.9 &  74.4 & 103.3 & -1.225 & -0.678 & -1.430 & -0.784 &  0.553 &  0.548 &  0.500 \\
            7  & [46.8, 167.3] & 162.4 &  64.9 & 171.4 &  1.348 & -0.467 &  1.596 & -0.565 & -0.347 & -0.354 & -0.311 \\
            8  & [44.6, 198.8] & 159.6 &  63.8 & 28.4  &  0.846 &  1.147 &  1.038 &  1.393 &  1.356 &  1.342 &  1.530 \\
            9  & [49.1, 147.7] & 145.8 &  58.3 & 17.9  &  1.136 &  0.711 &  1.381 &  0.876 &  0.626 &  0.634 &  0.724 \\
            10 & [69.8, 174.7] & 139.5 &  55.8 & 112.5 & -1.040 & -0.922 & -1.540 & -1.376 &  0.887 &  0.893 &  0.999 \\
            \bottomrule
         \end{tabular}
         \tiny}
         \label{tab:actual polarized}
    \end{table}
    
\begin{figure}[htbp]
        \centering
        \includegraphics[width=0.3\textwidth]{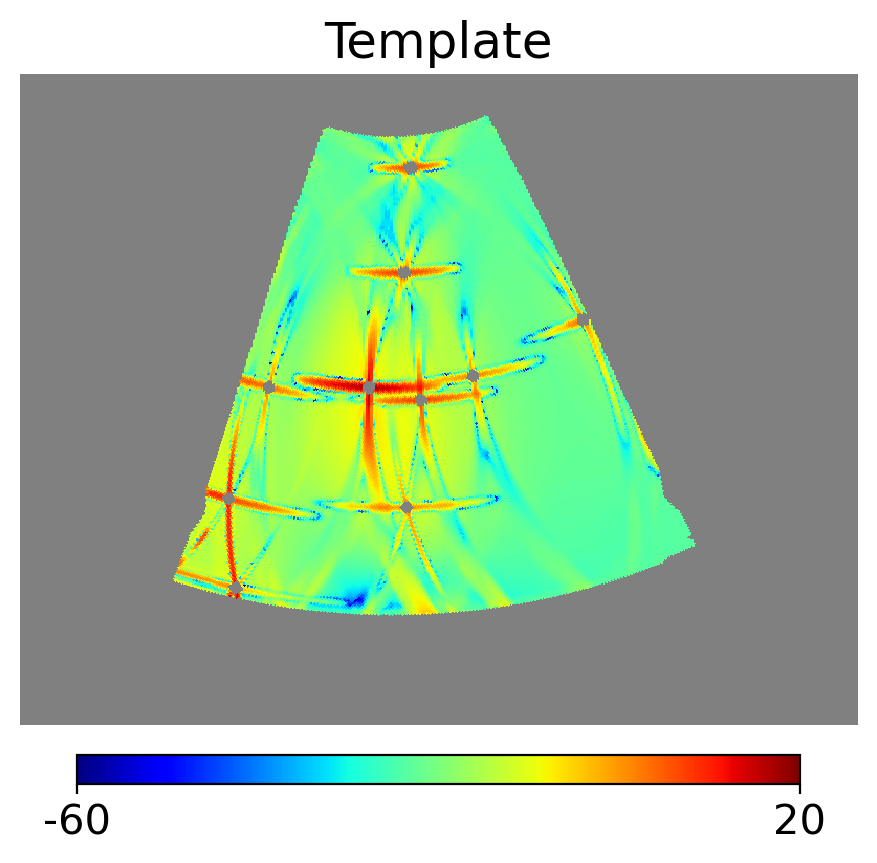}
        \includegraphics[width=0.3\textwidth]{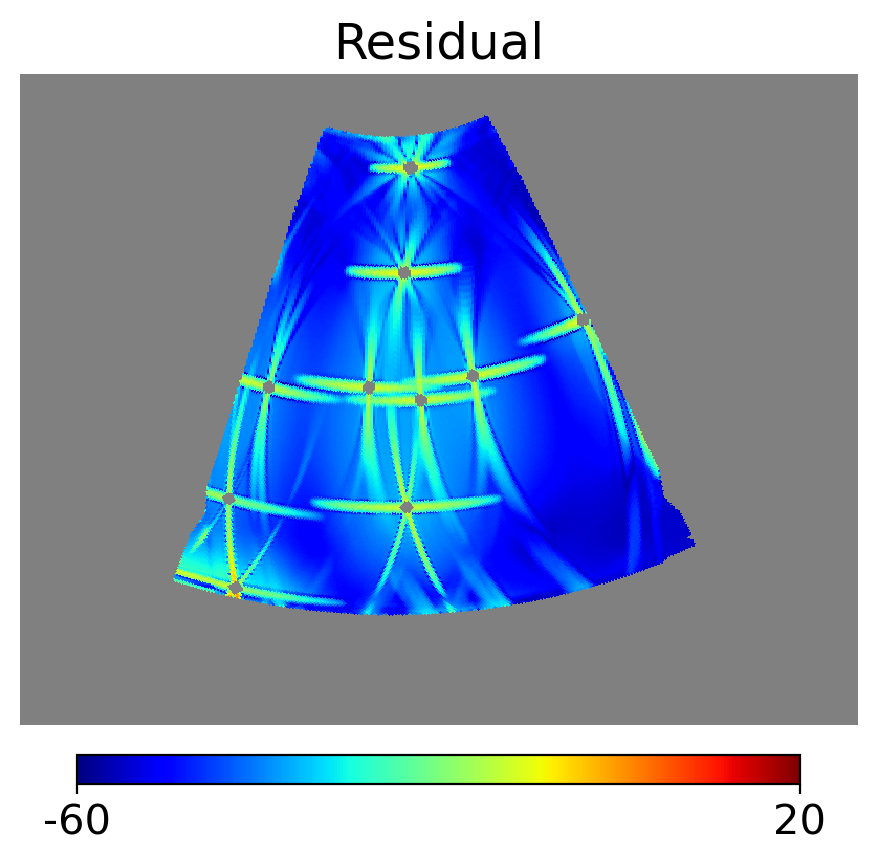}
        \includegraphics[width=0.3\textwidth]{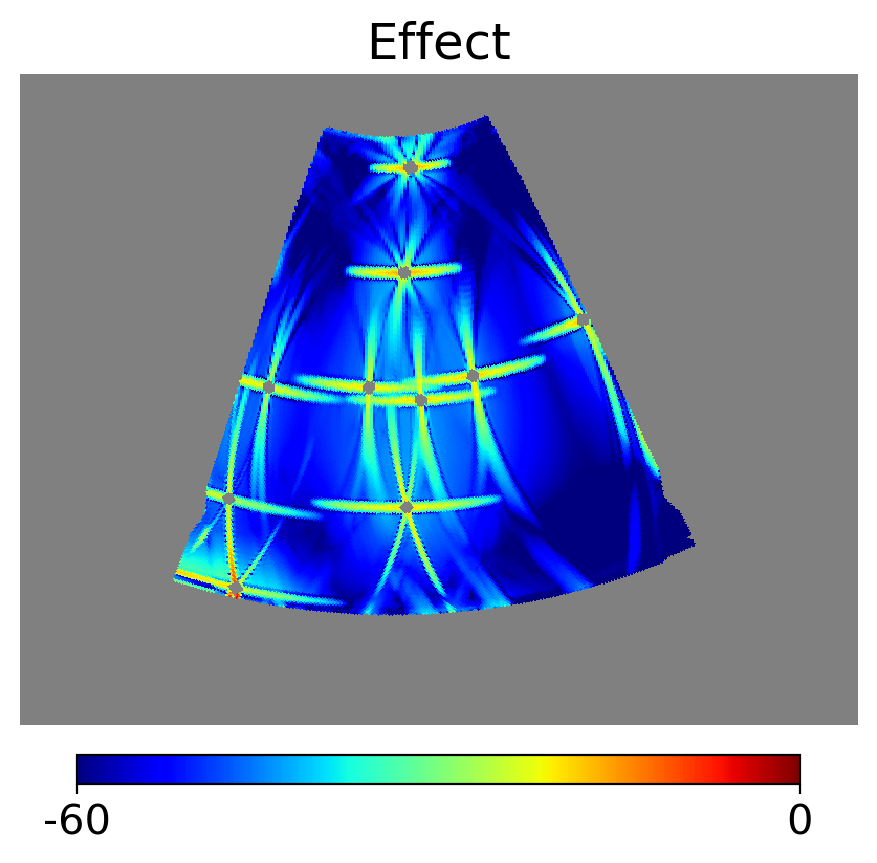}
        \includegraphics[width=0.3\textwidth]{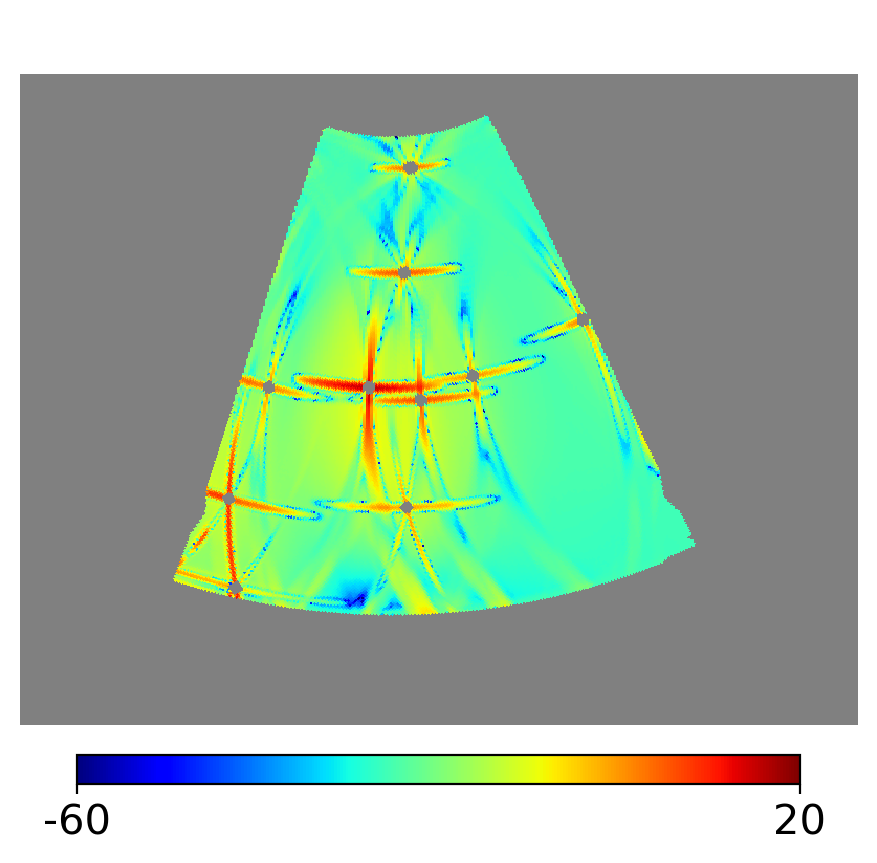}
        \includegraphics[width=0.3\textwidth]{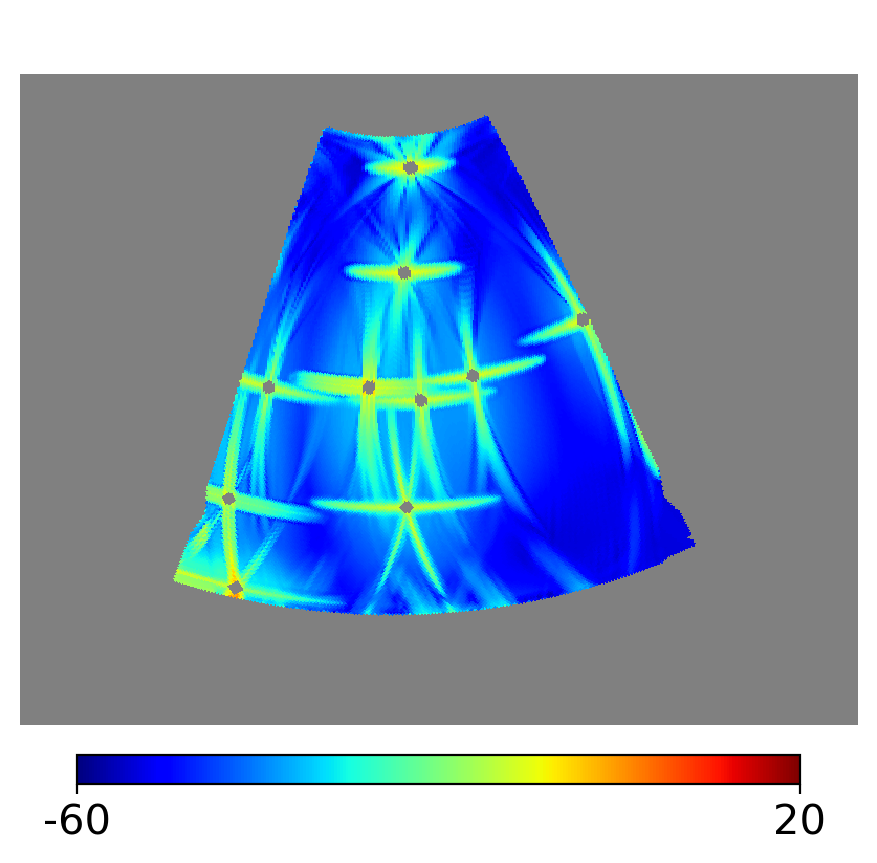}
        \includegraphics[width=0.3\textwidth]{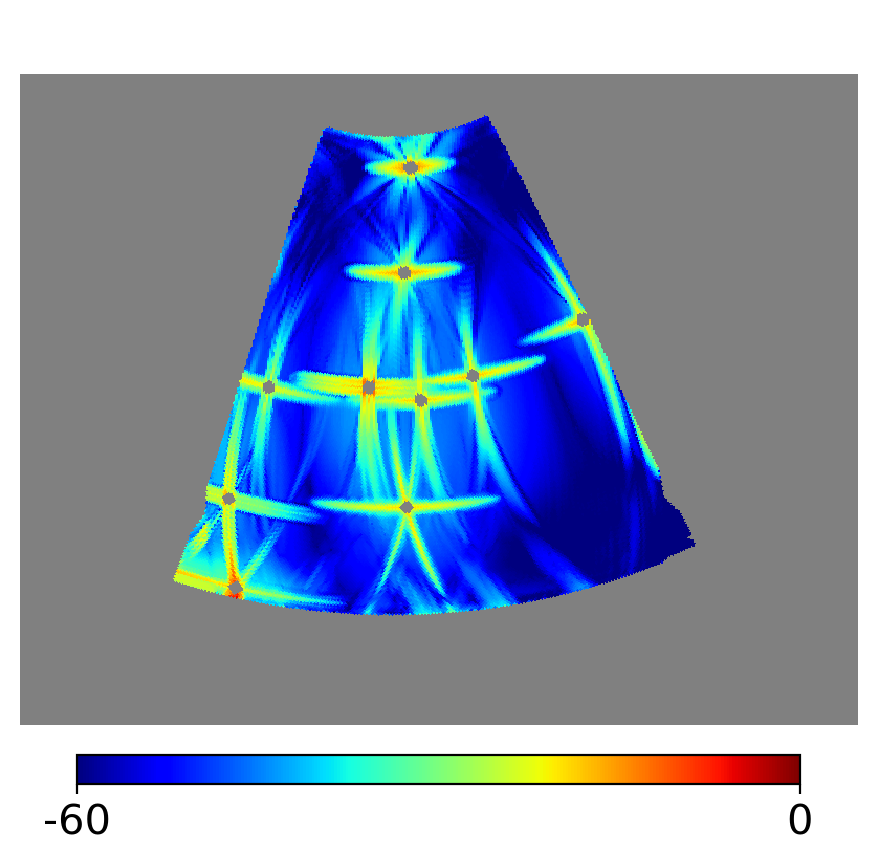}
        \caption{Same as in Fig.~\ref{fig:single sky map}, but for the case of actual point sources with unknown polarization direction, where the residuals are estimated using Eq.~\ref{eq:res1}.}
        \label{fig:actual sky map}
   \end{figure}

   We again build the sky map showing the polarization dB value of the templates, residuals and dB effects for actual point sources (Fig. \ref{fig:actual sky map}). The results are similar to Fig.~\ref{fig:multi sky map}, as detailed in Tab. \ref{tab:actual std}.
   
   \begin{table}[!h]
        \caption{Comparison of the standard deviation of $\sigma_*=10^{-2}$ $\mu\rm K$ of filtered CMB, foreground, noise, and actual 10 brightest point sources simulation residuals with unknown polarization direction in the pixel domain.}
        \centering
        \begin{tabular}{ccccc}
           \toprule
            $\sigma_{d'_c}/\sigma_*$ & $\sigma_{d'_f}/\sigma_*$ & $\sigma_{n'}/\sigma_*$ & $\sigma_{\delta_0}/\sigma_*$ & $\sigma_{\delta_1}/\sigma_*$ \\
           \midrule
            49.826 & 3.455 & 15.439 & 0.881  & 1.093 \\
           \bottomrule
        \end{tabular}
        \label{tab:actual std}
   \end{table}
    
\begin{figure}[htbp]
        \centering
        \includegraphics[width=0.45\textwidth]{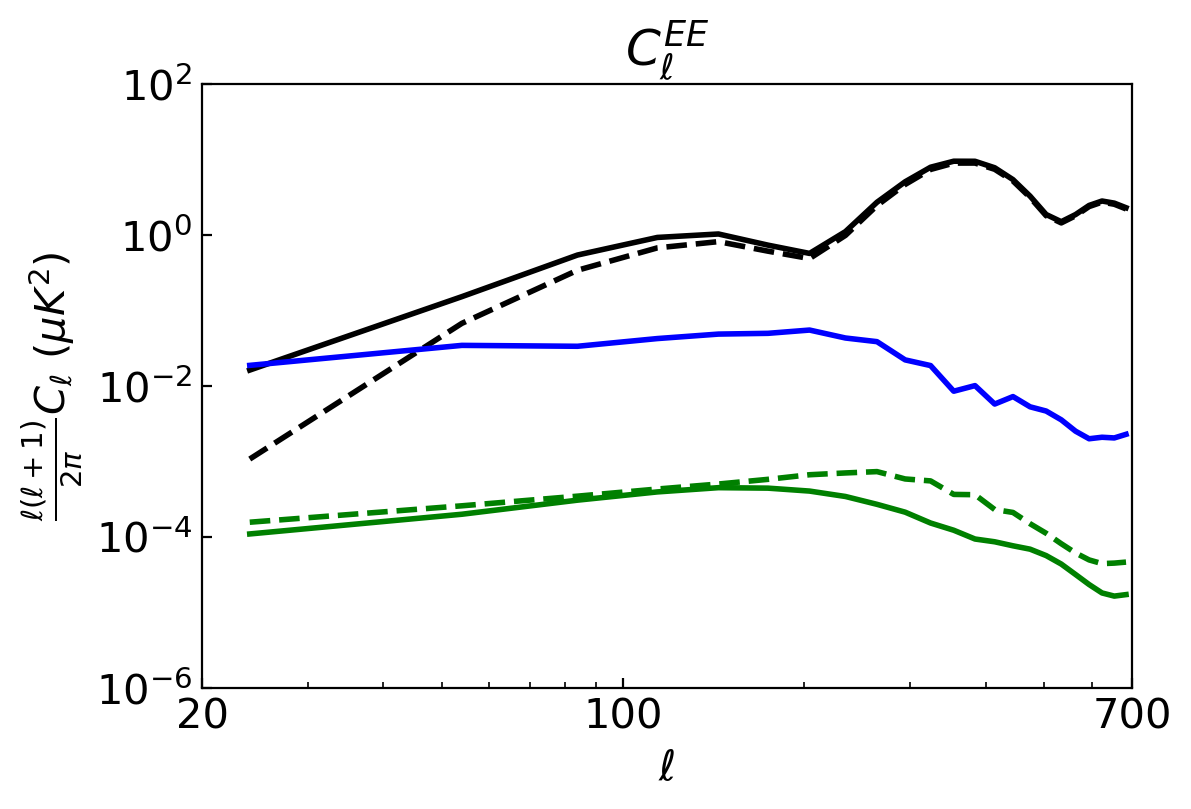}
        \includegraphics[width=0.45\textwidth]{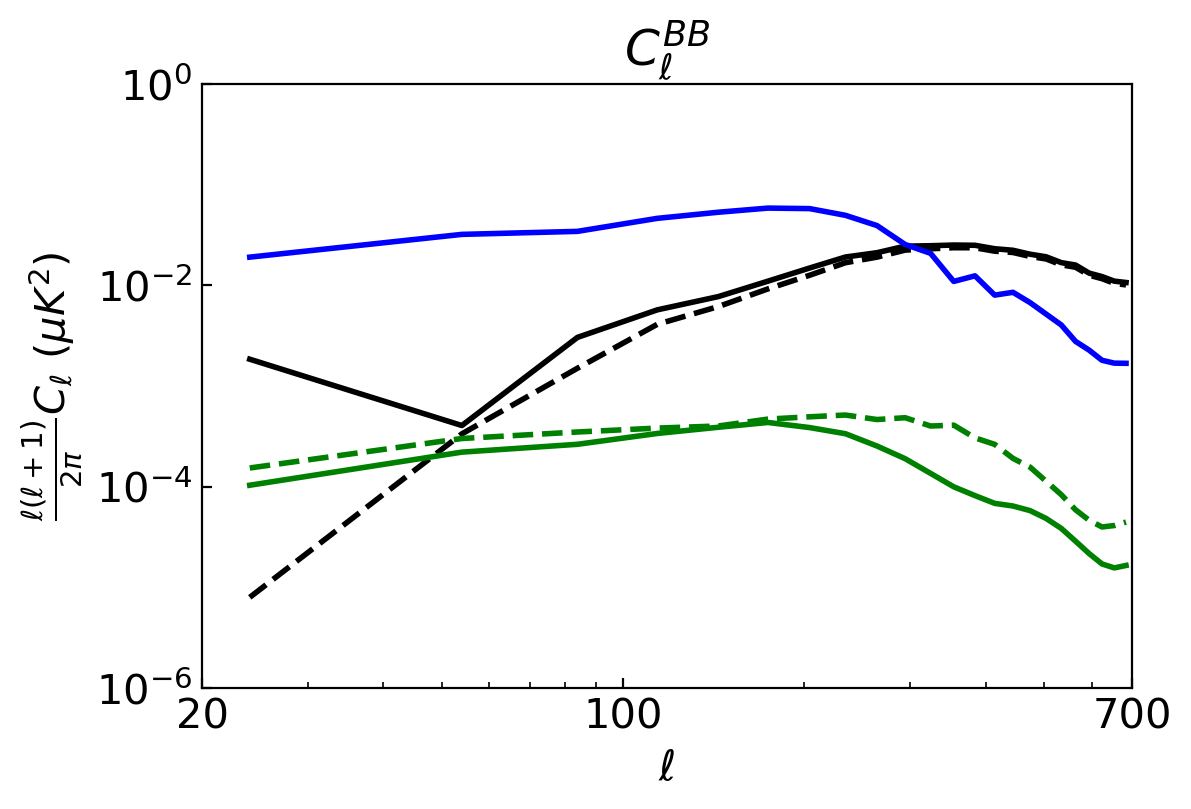}
        \caption{Same as in Fig.~\ref{fig:single cl}, but for
        the case of the actual point sources with unknown polarization directions.}
        \label{fig:actual cl}
    \end{figure}

    In Fig. \ref{fig:actual cl}, with our correction method, the angular power spectrum of residual is again substantially smaller than that of the true leakage and the expected CMB signal, demonstrating the effectiveness of our method with realistic point sources.

\section{Discussion}\label{sec:disscuss}

In this work, we have introduced a novel "template-fitting" method (Sect. \ref{sec:methods}) for removing the point source leakage due to time-order data filtering. The key ingredient of this method is to construct several templates of the leakage for each point source in pixel domain and then to remove this leakage contamination by fitting these templates. Several tests for single, multi and realistic point source simulations (Sect. \ref{sec:examples}) are present to demonstrate the effectiveness of our method. The leakage after template fitting is typically star-like, and  can be reduced by 1-2 orders of magnitude in the pixel domain and by 3-4 orders of magnitude in angular power spectrum. The performance of our method is robust in all simulations. According to the calculation of the angular power spectrum of the
residuals (see Figs.~\ref{fig:single cl},~\ref{fig:multi
cl},~\ref{fig:actual cl}), we can see that the residual $BB$ spectrum is about two orders of magnitudes lower than the theoretical prediction for the primordial gravitational waves
 with $r\sim10^{-2}$. Therefore, our method can easily satisfy the detection requirement for small $r$ down to $r\sim 10^{-3}$.

Using the matrix-based pipeline, the application of our template fitting method in this study becomes very practical. However, it is preferable to utilize a conventional non-matrix pipeline to execute our approach, which is because the construction of a full matrix for high-resolution map is impossible (scaled as $N^2_{\rm pix}$), due to the limitations in memory and storage. Fortunately, the standard non-matrix pipeline can be used with our method as well, as mentioned in Sect.~\ref{sub:realistic template}. In brief, all that needs to be done is to mask the map created by the standard pipeline with a point source mask, and then feed the masked sky map back into the pipeline to obtain the realistic templates. This fact allows our method to be functional even at high resolution, which is very useful in practice.

 \Ack{This work is supported by the National Key R\&D Program of China (2018YFA0404504, 2018YFA0404601, 2020YFC2201600, 2021YFC2203100, 2021YFC2203104), the Ministry of Science and Technology of China (2020SKA0110402, 2020SKA0110100), National Science Foundation of China (11890691, 11621303, 11653003), the China Manned Space Project with No. CMS-CSST-2021 (B01 \& A02), the 111 project No. B20019, and the CAS Interdisciplinary Innovation Team (JCTD-2019-05) and the Anhui project Z010118169.}




\providecommand{\href}[2]{#2}\begingroup\raggedright\endgroup

\end{document}